\documentclass[12pt,preprint]{aastex}
\def\trademark{{\hbox{\tiny TM}}}
\def\dof{{\rm dof}}
\def\min{{\rm min}}

\begin{document}
\title{The Most Precise Extra-Galactic Black-Hole Mass Measurement}

\author{Andrew Gould}
\affil{Department of Astronomy, Ohio State University,
140 W.\ 18th Ave., Columbus, OH 43210, USA; 
gould@astronomy.ohio-state.edu}

\begin{abstract}
I use archival data to measure the mass of the central black hole
in NGC 4526, $M=4.70\pm 0.14 \times 10^8\,M_\odot$.  This 3\%
error bar is the most precise for an extra-galactic black hole
and is close to the precision obtained for Sgr A* in the Milky Way.
The factor 7 improvement over the previous measurement 
is entirely due to
correction of a mathematical error, an error that I suggest may be
common among astronomers.

\end{abstract}


\section{{Introduction}
\label{sec:intro}}

\citet{davis13} have reported a mass measurement of the central black hole
in NGC 4526 of $M=4.5^{+4.2}_{-3.0}\times 10^8\,M_\odot$ ($3\,\sigma$)
based on interferometric CO emission measurements from its central molecular
ring.  They rightly regard their measurement as extremely important because
it opens the way to mass production of black hole mass 
measurements, particularly in the era of ALMA.

However, due to a mathematical error, these authors have vastly 
overstated their measurement errors and so vastly
underestimated the power of their technique.  In brief, they evaluated
their likelihood contours using $\chi^2/\dof$ rather than $\chi^2$ itself,
where dof equals ``degrees of freedom''.

Here I correct this error, thereby obtaining the most precise mass 
measurement of an extra-galactic black hole.  I argue that the CO
technique can be much more efficient than previously realized.
I also suggest that the substitution of $\chi^2/\dof$ for $\chi^2$
may be common among astronomers.

\section{{Data}
\label{sec:data}}

I used a Xerox$^\trademark$ machine to enlarge Figure 2 from \citet{davis13}
and measured the horizontal extent of the ``$1\,\sigma$'' error ellipse,
finding lower and upper boundaries at 3.65 and $5.75\times 10^8\,M_\odot$,
respectively.  I therefore adopt $4.70\pm 1.05\times 10^8\,M_\odot$ as the
``$1\,\sigma$'' mass estimate conveyed by this figure.  I note that the nearest
grid point evaluated by \citet{davis13} to my adopted central value is
at $4.5 \times 10^8\,M_\odot$, which is indeed the grid point with the lowest
$\chi^2$ reported by those authors.

\section{{Analysis}
\label{sec:analysis}}

The contours of Figure 2 of \citet{davis13} reflect $\chi^2/\dof$.  To
recover $\chi^2$, one should multiply by dof.  There are 68 data points
and two parameters, hence $\dof=68-2=66$.  However, \citet{davis13} report
$\chi^2_\min = 86.36$ and $\chi^2_\min/\dof=1.27$.  Hence, they are in fact
computing $\chi^2/\dof=\chi^2/68$.  We should therefore multiply by 68 rather
than 66.   

Before proceeding, we must evaluate $\chi^2$ at the actual minimum
of the error ellipse rather than the nearest grid point.  While the
nearest grid point is extremely close to the error-ellipse axis in the
ordinate direction, it is displaced from the center by 
$(4.70-4.50)/1.05 = 0.19$ ``$\sigma$'' in the abscissa
direction.  Hence, at the ellipse center,
the quantity that \citet{davis13} are calling ``$\chi^2/\dof$'' should
be lower by $0.19^2= 0.036$, implying $\chi^2_\min = 83.91$.

Before evaluating the true error bar, we must take note of the fact
that $\chi^2_\min/\dof = 83.91/66 = 1.27$ is greater than unity.
That is, the expected $(1\,\sigma)$ range of this parameter is
$1\pm (2/\dof)^{1/2}=1\pm 0.174$, so that the actual value is too high by 
$1.6\sigma$.  There are three possible reasons for this high value:
1) normal statistical fluctuations, 2) underestimated errors, 
3) inadequacy of the model.  To be conservative, I adopt explanation (2)
and assume that the measurement errors have been slightly underestimated,
i.e., by a factor $\sqrt{1.27}=1.127$.

If there were no such correction, then the true error would be smaller
than the reported one by $\sqrt{68}$, because \citet{davis13} divided
their $\chi^2$ values by 68.  However, taking account of the slightly
underestimated errors, the true correction factor is $\sqrt{68/1.27}=7.32$.
Therefore, I finally derive a mass estimate for the central black hole
of NGC 4526 of
\begin{equation}
M = 4.70\pm 0.14\times 10^8\,M_\odot.
\label{eqn:bhmass}
\end{equation}

\section{{Discussion}
\label{sec:discuss}}

\subsection{{Most Precise Mass Measurement}
\label{sec:precise}}

The first point to note is that the statistical
error bar in Equation~(\ref{eqn:bhmass}) is only 3\%.  
This is not quite at the level the statistical errors for the Milky Way's
own central black hole $4.4\times 10^6\,M_\odot$ \citep{ghez08} or
$4.28\pm 0.07\times 10^6\,M_\odot$ \citep{gillessen09} at fixed $R_0=8.3\,$kpc.
However, it is more precise than any extra-galactic black hole measurement.
For example, the spectacular nearby mega-maser NGC 4258 yields only a 6\%
statistical-error black hole mass measurement 
$3.3\pm 0.2 \times 10^7\,M_\odot$ at fixed distance 7.28 Mpc \citep{stopis09}.

\subsection{{Precision vs.\ Accuracy}
\label{sec:accuracy}}

Normally in astronomy, one is more concerned about accuracy than precision,
both in the sense that one is finally interested in how close the
reported measurements are to the true quantities and in the sense that
estimating accuracy usually requires quite a bit more effort than estimating
precision.

In the present case, 
the mass estimate scales with the assumed distance of NGC 4526,
which is fixed at 16.4 Mpc in the analysis of \citet{davis13}, who cite the
surface brightness fluctuation distance measurement by \citet{tonry01}.  
The latter work
quotes a 10\% distance error (and a distance of 16.9 Mpc).  In principle,
one might hope to do better, but even for galaxies in the Hubble flow, there
is an error floor due to uncertainty in the Hubble constant.  In addition,
there are many other sources of systematic error, such as assumptions within
the modeling about the geometrical structure of both the CO gas and the
stars.  In brief, there is no immediate hope of achieving accuracies at the
level of this 3\% precision.

Nevertheless, I would argue that precision is indeed the key parameter
of interest.  The point is that \citet{davis13} could have achieved
10\% precision in only about 1/10 of their observing time.  While they
do not specify exactly how much observing time they used, they do say
that similar precision could have been achieved by 5 hrs of ALMA integration
for an object at 75 Mpc.

Now, 5 hrs of ALMA observing time cannot simply be scaled to 30 minutes
at a cost of a factor 3 in precision because ALMA relies on the rotation
of the Earth to fill in the UV plane.  However, with some ingenuity, one
could organize observations of, say, six objects, and sequentially rotate
through them over 5 hours.  That is, a vast program of precision black-hole
mass measurement could be undertaken at relatively little cost.

\subsection{{$\chi^2$ vs.\ $\chi^2/\dof$}
\label{sec:chi2}}

This is the second paper in as many months for which I noticed that
the authors calculated confidence contours based on changes in
$\chi^2/\dof$ rather than $\chi^2$.  Given that it is not often
that I read a paper in which there is sufficient information to determine
which parameter was used, I think that this error might be quite common.
More striking than the errors themselves is the fact that I am the only one
who seems to notice them.  The two papers had between them 15 co-authors
and at least 4 referees.  In both cases, the effect of the error was
to radically degrade the precision of the reported measurement and
so should have jumped out to anyone with a stake in the paper or its
results.

For the record, for Gaussian-distributed errors (which, from the central
limit theorem, usually applies to the case of many dof), likelihood scales
as $L\sim\exp(-\Delta\chi^2/2)$ independent of the number of dof. 
It is therefore from $\chi^2$ itself that one calculates the relative
likelihood of models.  

On the other hand, $\chi^2/\dof$ is useful
primarily to check the quality of the data and the completeness of the
model space.  If the model space covers the system being modeled, and
the error bars are correctly estimated, then $\chi^2/\dof$ should be
about unity.  Strictly speaking, $\chi^2$ will under these conditions
be distributed as a ``$\chi^2$ distribution'', which has a mean and
standard deviation of $\dof\pm\sqrt{2\dof}$.  Hence, if the observed
value of $\chi^2$ lies well outside these bounds, it demonstrates that
either 1) the errors have been misestimated and/or 2) the model space is
inadequate.  See \citet{gould03} for a more detailed discussion.

%\begin{equation}
%\label{eqn:}
%\end{equation}

\acknowledgments

I thank Chris Kochanek for helpful discussions.
This work was supported by NSF grant AST 1103471.

%\begin{figure}
%\plotone{beta.eps}
%\caption{\label{fig:}
%}
%\end{figure}

\end{document}